\documentclass[12pt]{article}
\usepackage{amsfonts,amsmath,amssymb}
\DeclareMathOperator\Res{Res}
\DeclareMathOperator\Tr{Tr}

\begin{document}

\title 
{Richardson-Gaudin Algebras and the Exact Solutions of the Proton-Neutron Pairing}
\author
{Vesselin G. Gueorguiev$^1$ and Jorge Dukelsky$^\dag$ \\ \\
\it \small $^{1}$Institute for Nuclear Research and Nuclear Energy,\\
\it \small Bulgarian Academy of Science, Sofia, Bulgaria\\
\it \small $^\dag$ Instituto de Estructura de la Materia, CSIC,\\
\it \small Serrano 123, 28006 Madrid, Spain\\}
\date{\small Contribution to the XII International Conference on \\
\small Geometry, Integrability and Quantization, \\
\small June 04 - 09, 2010, Varna, Bulgaria}
\maketitle

\begin{abstract}
Many exactly solvable models are based on Lie algebras. The pairing interaction is important in nuclear physics and its exact solution for identical particles in non-degenerate single-particle levels was first given by Richardson in 1963. His solution and its generalization to Richardson-Gaudin quasi-exactly solvable models have attracted the attention of many contemporary researchers and resulted in the exact solution of the isovector pn-pairing within the $ \frak{so}(5)$ RG-model and the equal strength spin-isospin pn-pairing within the $ \frak{so}(8)$ RG-model. Basic properties of the RG-models are summarized and possible applications to nuclear physics are emphasized.
\end{abstract}

\stepcounter{footnote}
\footnotetext{On leave of absence from the Institute for Nuclear Research and Nuclear Energy, Bulgarian Academy of Science, Sofia, Bulgaria. For current mailing address look up the APS members directory or send e-mail to vesselin at mailaps.org.}

\section{Introduction}

Symmetry is one of the most important paradigms in modern physics. Any Lie group and its algebra have a naturally defined action on a product of spaces (representations). Thus, they are very suitable for a multi-particle system with an underling symmetry. Usually, this means that the relevant operators are well defined for one particle as well as for any number of particles. This allows one to find exact solutions to a problem, with a given underlining symmetry, by referring to the relevant representation theory of the symmetry group at place. As a result many exactly solvable models are build using Lie algebra representation theory. A well know examples are the $\frak{so}(3)$ and 
$ \frak{su}(2)$ rotational symmetry, the Elliott's $ \frak{u}(3)$ symmetry model \cite{Elliott1, Elliott2, Elliott3, Elliott4}, the Wigner's $\frak{su}(4)$ spin-isospin symmetry \cite{Wigner}, and many more that play a major role in nuclear physics. For example, the $\frak{so}(8)$ and $\frak{sp}(6)$ Ginnocchio models, the Fermion Dynamical Symmetry Models (FDSM), and the three dynamical symmetries of the Interacting Boson Model (IBM). 

A nuclear many-body system near equilibrium can be viewed as subject to a mean field Harmonic Oscillator (HO) potential:
\[H_{0}=\frac{\vec{p}^2}{2m}+\frac{1}{2}k^2 \vec{x}^2.\]
Since the symmetry group of the  3-dimensional HO is $\frak{u}(3)$ \cite{Elliott1} one can easily see its relevance in the description of nuclei. It is well know that one can understand the magic numbers and the shell structure of nuclei within the 3-dimensional HO approximation  \cite{Haxel&Jensen}. Using the HO single-particle states  one can write a general  Hamiltonian with one- and two-body terms:
\[ H=\sum_{i}\varepsilon_{i}a^{+}_{i}a_{i}+\frac{1}{4}\sum_{i,j,k,l}V_{ij,kl}a^{+}_{i}a^{+}_{j}a_{k}a_{l}.\]
Here, $a_{i}$ and $a^{+}_{j}$ are fermion annihilation and creation operators, $\varepsilon_{i}$ single-particle energies, and $V_{ij,kl}=\left< ij|V|kl\right>$ two-body interaction matrix elements.

The simplest extension of the HO Hamiltonian is to add quadruple interaction terms $Q\cdot Q$ and/or spin-orbit interaction $L.S$. This has been well studied by Elliott and his collaborators \cite{Elliott1, Elliott2, Elliott3, Elliott4}. 

Another important interaction  is the pairing interaction in nuclei:
\begin{equation}
H_{P}=\sum_{i}2\varepsilon_{i}n_{i}-
g\sum_{i,j}a_{\uparrow ,i}^{+}a_{\downarrow ,i}^{+}a_{\uparrow,j}a_{\downarrow ,j}.
\label{H_P}
\end{equation}
Here $n_{i}$ is the number operator for pairs. The exact solution of the pairing interaction 
between identical particles in non-degenerate single particle levels was first given by 
Richardson \cite{Ric63, Ric66}. His solution and its generalization to Richardson-Gaudin 
quasi-exactly solvable models (RG-models) have attracted the attention of various contemporary 
researchers - resulting in the exact solution of the isovector proton-neutron pairing in nuclei within 
the $ \frak{so}(5)$ RG-model \cite{so5pairing} -` to be discussed in Sec. \ref{so5model} 
and the equal strength spin-isospin proton-neutron pairing within the $ \frak{so}(8)$ 
RG-model \cite{so8pairing} to be summarized in  in Sec. \ref{so8model}. 
Basic properties of the integrable RG-models are summarized in Sec. \ref{RGmodel} and their possible applications 
to variety of nuclear physics models are emphasized in Sec. \ref{nuclearmodels}. In the next section we briefly discuss few dynamical symmetry models of importance to nuclear systems.

\section{Some Exactly Solvable Nuclear Models}

A quantum system has a dynamical symmetry if the Hamiltonian can be
expressed as a function of the Casimir operators of a subgroup chain. A
typical example of a rank two dynamical symmetry is the Elliott's $\frak{su}(3)$
model that is used in the description of deformed nuclei \cite{Elliott1, Elliott2, Elliott3, Elliott4}. 
\begin{eqnarray}
H_{def}=\varepsilon N+\chi Q\cdot Q.\label{Hdef}
\end{eqnarray}
This Hamiltonian can be rewritten as a linear combinations of the Casimir operator of the $\frak{su}(3)$
Lie algebra $C^{\frak{su}(3)}_{2}=(Q\cdot Q+3L^2)/4$ involving the quadrupole-quadrupole interaction $Q\cdot Q$  and the Casimir operator $L^2$ of the $\frak{so}(3)$ subgroup of the angular momentum. 
Thus we have a group chain reduction $\frak{so}(3)\subset \frak{su}(3)$ that provides exact solution to our initial Hamiltonian (\ref{Hdef}):
\[
H_{\frak{so}(3)\subset \frak{su}(3)}=\varepsilon N+ \frac{1}{2 \frak{J}} L^2+a C^{\frak{su}(3)}_{2}.
\]
Here $N$ counts the number of particles with energy $\varepsilon$ of a particular HO shell and $L^2$ lifts the $l$-degeneracy of a particular harmonic oscillator shell. 

A common feature of the dynamical symmetry nuclear models, is that they are
all defined for a degenerate single particle levels. Since single particle energy 
splitting breaks the dynamical symmetry, it is usually expected that this will 
prevent the model to be exactly solvable. For example,  
spin-orbit interaction $l\cdot s$ lifts the total angular momentum degeneracy 
$j=l+1/2$ and $j=(l+1)-1/2$ and destroys the $\frak{su}(3)$ symmetry \cite{VGG01}.  
Although, the single particle energy splitting breaks the dynamical symmetry it may still preserve the exact solvability. The pairing model with non-degenerate single particle levels,
whose exact solution has been found by Richardson, represents a unique example of an exactly solvable model with these characteristics \cite{Ric63, Ric66}. The model is exactly-solvable due to the special extension of the relevant dynamical symmetry algebra to a spectral Gaudin algebra.

\section{Spectral Lie Algebras}\label{RGmodel} 

A Gaudin algebra $\mathcal{G}\left( \frak{g}\right) $  is an infinite dimensional extension of a Lie algebra $\frak{g}$ that associates to any generator $X^{\alpha } \in\frak{g}$ a parameter dependent generator $X^{\alpha }\left( \lambda \right)  \in \mathcal{G} \left(\frak{g}\right) $ satisfying the following commutation relations \cite{Ushveridize}:

\begin{equation}
\left[ X^{\alpha }\left( \lambda \right) ,X^{\beta }\left( \mu \right)
\right] =\sum_{\gamma }\Gamma _{\gamma }^{\alpha \beta }\frac{X^{\gamma
}\left( \lambda \right) -X^{\gamma }\left( \mu \right) }{\lambda -\mu }.
\label{Gau}
\end{equation}
Here $\Gamma _{\gamma }^{\alpha \beta }$ are the structure constants of the Lie algebra $\frak{g}$; 
$\lambda$ and $\mu$ are complex spectral parameters. One can form Hermitian operators by using the dot product defined via the $\frak{g}$-invariant metric tensor $g^{\alpha \beta }\sim \Tr(ad(X^{\alpha}) ad(X^{\beta})) $:
\begin{equation}
K\left( \lambda \right) =X(\lambda)\cdot X(\lambda) \label{Inte}.
\end{equation}
$K\left( \lambda \right) $ are not Casimir operators because they do not commute with all generators of $\mathcal{G}\left(\frak{g}\right) $. However, these operators commute among themselves 
 
\begin{equation}
\left[ K\left( \lambda \right) ,K\left( \mu \right) \right] =0 \label{Com}.
\end{equation}
This implies that the system is integrable and $K(\lambda)$ are the integrals of motion.

There is a unique rational realization\footnote{There are non-rational Gaudin algebra realizations but we will not consider them here.} of the generators $X^{\alpha }\left(
\lambda \right) $ in terms of the generators of $L$ copies of the algebra 
$\frak{g}$ given by the following expression: 
\begin{equation}
X^{\alpha }\left( \lambda \right) =\sum_{i=1}^{L}\frac{1}{z_{i}-\lambda }
X_{i}^{\alpha }+\rho ^{\alpha }, \label{Rep}
\end{equation}
The $z_{i}$ are $L$ arbitrary numbers, which will ultimately be
related to the single particle energies. Here we deviate from \cite{Ushveridize} by
introducing the set of arbitrary parameters $\rho ^{\alpha }$. However, most
of the expressions related to the Gaudin algebra (\ref{Gau}) derived for the
case $\rho ^{\alpha }=0$ are still valid because $X^{\alpha }\left( \lambda
\right) \rightarrow X^{\alpha }\left( \lambda \right) +\rho ^{\alpha }$ is
an algebra isomorphism. The shift of the elements of $\mathcal{G}$ by the
$\rho$ parameters is a key to the symmetry breaking in the model.

\subsection{Richardson-Gaudin Operators}

For the realization (\ref{Rep}) of the Gaudin algebra, the integrals of motion are: 
\[
K\left( \lambda \right) =\rho \cdot \rho +\sum_{i}\frac{C_{2}^{(i)}}{\left( z_{i}-\lambda \right) ^{2}}+2\sum_{i}\frac{R_{i}}{z_{i}-\lambda }.
\]
$C_{2}^{(i)}$ is the second Casimir operator of the $i$-th copy of 
$\frak{g}$. Thus the first two terms are constants for states built on a
tensor product of irreducible representations of $\frak{g}$. The $R_{i}$
are the Richardson-Gaudin operators \cite{Gaudin-HdeS, R-G-models} and are one-half of the residue of $K(\lambda)$ at $\lambda =z_{i}$ : 
\begin{equation}
R_{i}=\sum_{j(\neq i)}\frac{X_{i}\cdot X_{j}}{z_{i}-z_{j}}+\xi _{i},
\quad \xi _{i}=\rho\cdot X_{i} \label{Rop}.
\end{equation}
By taking the residues of (\ref{Com}) at $\lambda =z_{i}$ and $\mu =z_{j}$
one can see that the $R_{i}$ operators commute among themselves. Therefore,
they define a new set of integrals of motion. Thus any function of the 
$R_{i} $ operators can be used as a model Hamiltonian for this integrable
system. In particular, any linear combination of the $R_{i}$ operators is at
most quadratic in the generators.

For a singular semi-simple algebras, the eigenvalues of the $R_{i}$
operators can be obtained from the eigenvalues $k(\lambda)$ of the 
$K(\lambda)$ operator given in \cite{Ushveridize} by taking the appropriate residue
(1/2 $\Res(k(\lambda),\lambda =z_{i})$): 
\begin{equation}
r_{i}= \Lambda _{i}\cdot\rho+\sum_{j(\neq
i)}\frac{\Lambda _{i}\cdot\Lambda _{j}}{z_{i}-z_{j}}
+\sum_{a=1}^{r}\sum_{\alpha =1}^{M^{a}}\frac{\Lambda _{i}\cdot\pi^{a}}{
z_{i}-e_{a,\alpha }}. \label{regv}
\end{equation}
Here $e_{a,\alpha }$ are solutions of the generalized
Richardson equations \cite{Ushveridize, Links-JPA35, Asorey-et-al}: 
\begin{equation}
\sum_{b=1}^{r}\left. \sum_{\beta =1}^{M^{b}}\right. ^{\prime }\frac{
\pi ^{b}\cdot\pi ^{a}}{e_{b,\beta }-e_{a,\alpha }}-\sum_{i=1}^{L}\frac{
 \Lambda _{i}\cdot\pi ^{a} }{z_{i}-e_{a,\alpha }}
= \xi \cdot\pi ^{a}.
 \label{gReq}
\end{equation}
$L$ is the number of copies of the algebra $\frak{g}$ and $\Lambda _{i}$ is the weight for the $i$-th copy. 
The $\Lambda _{i}\cdot\pi ^{a}$ are actually the eigenvalues of the generators 
$H^{a}=\pi _{s}^{a}h^{s}$ at the lowest/highest weigh state of the 
$i$-th copy of $\frak{g}$. The $\pi _{s}^{a}$ are the components of the
positive simple roots\footnote{$E^{a}$ is a simple root vector if it cannot be witten as a commutator of any other two positive root vectors.} $\pi ^{a}$ of the Lie algebra $\frak{g}$ in the Cartan-Weyl basis $([h_{s},E^{a}]=\pi _{s}^{a}E^{a})$, and 
$\xi \cdot\pi ^{a} =\pi _{s}^{a}\rho ^{s},$ where $\rho ^{s}$
are the components of the symmetry breaking one-body operator 
$\xi =\rho ^{s}h_{s}$ along the Cartan generators $h_{s}$. The rank of $\frak{g}$ is 
$r$ and $M^{a}$ are positive numbers related to the eigenvalues $m^{a}$ of 
$H^{a}$ at the desired eigenstate ($M^{a}=\sum_{i}\Lambda _{i}^{a}-m_{i}^{a}$).

\subsection{Symmetry Breaking}

Although one can use any function of $R_{i}$ as a Hamiltonian, there is a
particular liner combination of the Gaudin operators $R_{i}$ that results in
a simple expression which is linear in the spectral parameters $e_{a,\alpha
}.$ From this expression one can see that the breaking of the 
$\frak{g}$-symmetry is due to the $\xi $ terms:
\begin{equation}
H=\sum_{i}z_{i}R_{i}=\sum_{i}z_{i}\xi _{i}-\frac{1}{2}\left(
C_{2}-\sum_{i}C_{2}^{(i)}\right).
\label{Ham1}
\end{equation}
In general $z_{i}\neq z_{j},$ thus the term $\sum_{i}z_{i}\xi _{i}$ would
mix different irreps because it will not commute with the total Casimir
operator $C_{2}$ that is built from the generators $X=$ $\sum_{i}X_{i}.$ The
final symmetry of this Hamiltonian is determined by the set of generators
that commute with $\xi .$ In order to see that the eigenvalues of (\ref{Ham1}) are linear in the spectral parameters $e_{a,\alpha }$ one has to multiply
the generalized Richardson equations (\ref{gReq}) by $e_{a,\alpha }$ and sum
over all the indexes. After some manipulations one would observe that the last
term in (\ref{regv}) appears in the relevant expressions.

As we already discussed in the previous section, common feature of the dynamical symmetry nuclear models is that they are all defined for degenerate single particle levels. Single particle energy splitting breaks the dynamical symmetry but may still preserve the integrability. The pairing model with non-degenerate single particle levels, whose exact solution has been found by Richardson, along with the above discussed Richardson-Gaudin constructions are examples of exactly solvable models with such characteristics \cite{Ric63}.

Models based on fermion realization of the generators of the type $a_{\alpha }^{+}a_{\beta }^{+}$, such as the $\frak{sp}(2n)$ algebras, are naturally suitable for non-degenerate single particle systems. This is related to the fact that the
fermion realizations of the corresponding Cartan generators are related to
the fermion number operators. This is an easy observation if one looks at
the commutator $[X_{\alpha \beta }^{+},X_{\alpha \beta }^{-}]=n_{\alpha
}+n_{\beta }-1$ where the non-Cartan generators are in the form $X_{\alpha
\beta }^{+}=a_{\alpha }^{+}a_{\beta }^{+}$ and $(a_{\alpha })^{+}=a_{\alpha
}^{+}.$ Note that $n_{\alpha }$ and $n_{\beta }$ enter on an equal footing. In contrast models that are built on generators of the form $X_{\alpha \beta}^{+}=a_{\alpha }^{+}a_{\beta }$, such as $\frak{su}(n)$ algebras, result in Cartan generators of the form $n_{\alpha }-n_{\beta }.$

\section{Pairing in Nuclei}

Now we will direct our discussion towards the applications of the generalized Richardson-Gaudin models in nuclear physics. For this reason we shall specify $z_{i}$ to be related to the single particle energies $\varepsilon _{i}$ by the expression $z_{i}=2\varepsilon _{i}.$ $\rho^{\alpha }$ to be non-zero only for $X^{\alpha }$ that are elements of the
Cartan algebra of $\frak{g},$ that is $\rho ^{a}\neq 0$ and $a=1...r$. Then a ``pairing'' like Hamiltonian $H_{P}$ is obtained from (\ref{Ham1}) by considering $H_{P}=g H$ and its form is: 
\begin{eqnarray*}
H_{P} &=&\sum_{i}2\varepsilon _{i}\delta\cdot h_{i}-
g \sum_{\beta \in \Omega _{+}}\sum_{i\ne j}Y_{i}^{-\beta}Y_{\beta,j} \\
E_{P} &=&\sum_{a=1}^{r}\sum_{\alpha=1}^{M^{a}}e_{a,\alpha}\delta _{a},\quad 
g=\prod_{\rho^{a}\neq 0}\frac{1}{ \rho ^{a} },\quad \delta_{a}=g\rho_{a}.
\end{eqnarray*}
The $E_{P}$ are the eigenvalues of $H_{P}$ and $Y_{\beta }$ are the positive
root vectors with respect to the chosen Cartan algebra $\left\{
h^{a}\right\} $ of $\frak{g}.$ Now $g $ plays the row of a coupling
constant for the two-body interaction term $Y^{-\beta }Y_{\beta }.$ 

This is particularly clear in the case of the pairing model where $\frak{g}\Bbb{=}
\frak{su}(2)$ with generators $Y_{+}=a_{\uparrow }^{+}a_{\downarrow }^{+}$, 
$Y_{-}=a_{\uparrow }a_{\downarrow }$ and $h=a_{\uparrow }^{+}a_{\uparrow
}+a_{\downarrow }^{+}a_{\downarrow }.$ Since there is only one Cartan
generator $h$, there is only one $\rho=\frac{1}{g}$, and $\delta=1,$ and 
$E_{P}$ results in the usual expression of a sum over
the pair energies $e_{i},$ $E_{P}=\sum_{i=1}^{M}e_{i}$ for the standard
pairing Hamiltonian (\ref{H_P}).

\subsection{T=1 Proton-Neutron Pairing as $\frak{so}(5)$ RG-model}\label{so5model}

To better illustrate the current framework we will briefly discuss the $T=1$
pairing in proton-neutron systems which is related to an $\frak{so}(5)$ Lie algebra 
\cite{Links-JPA35}. In this case the one-level system is constructed from
various proton-proton, neutron-neutron, and proton-neutron pairs. By
choosing the Cartan generators to be the total particle number operator 
$h_{2}=1-(\hat{N}_{p}+\hat{N}_{n})/2$ and the third projection of the isospin 
$h_{1}=T_{0}=(\hat{N}_{p}-\hat{N}_{n})/2.$ We find the positive root 
vectors\footnote{Positiveness of a root vector is determined by the positiveness of the
corresponding eigenvalues: first with respect to $h_{r}$, 
if zero then one has to look at $h_{r-1}$ and so on.} of the algebra to be the
hard boson annihilation operators $\{b(\mu):\mu=1,2,3\}=$ $\{n_{\uparrow
}^{-}n_{\downarrow }^{-},(p_{\uparrow }^{-}n_{\downarrow }^{-}+n_{\uparrow
}^{-}p_{\downarrow }^{-})/\sqrt{2},p_{\uparrow }^{-}p_{\downarrow }^{-}\}$
plus the isospin rising operator $T_{+}=(p_{\downarrow }^{+}n_{\downarrow
}^{-}+p_{\uparrow }^{+}n_{\uparrow }^{-})/\sqrt{2}$. The simple root vectors
are $\left\{ b(3),T_{+}\right\} $ and $b(3)=p_{\uparrow }^{-}p_{\downarrow
}^{-}$ is the singular root vector \cite{Ushveridize}. In the chosen basis
the corresponding commutation relations are: 
\begin{eqnarray*}
\lbrack h_{2},h_{1}] &=& 0,\quad [b(\mu),b(\nu)]=0,\quad \mu,\nu=1,2,3\\
\lbrack h_{2},b(\mu)] &=& b(\mu), \quad \lbrack h_{2},T_{+}] =0,
\quad [h_{1},T_{+}]=T_{+}\\
\lbrack h_{1},b(1)] &=&b(1),\quad [h_{1},b(2)]=0,\quad [h_{1},b(3)]=-b(3) \\
\lbrack b(3),T_{+}] &=&b(2),\quad [T_{+},b(2)]=b(1),\quad [T_{+},b(1)]=0.
\end{eqnarray*}
By including the conjugated operators one closes the algebra $\frak{so}(5)$.
The isospin sub-algebra  $\frak{su}_{T}(2)\subset \frak{so}(5)$ is generated by the 
$\frak{so}(5)$ generators that are not related to the singular root vector $b(3)=p_{\uparrow }^{-}p_{\downarrow}^{-}.$ That is, $\frak{su}_{T}(2)$ is generated by $T_{+}$,  $T_{-}=(T_{+})^{+}$, and $[T_{+},T_{-}] = T_{0}= h_{1}$. Even more, all the given 
$\frak{so}(5)$ generators can be recognized as $\frak{su}_{T}(2)$ tensor operators.

Since $\frak{so}(5)$ is a rank two algebra, we have two types of spectral
parameters: $e_{1,\alpha }$ and $e_{2,\beta }$ which we will denote by 
$w_{\alpha }$ and $v_{\beta }$ in the following discussion. The upper bounds 
$M^{1}$ and $M^{2},$ for the indices $\alpha $ and $\beta $, are related to
the isospin $T$ and the total number of pairs $M$ via the expressions 
$M^{1}=M-T$ and $M^{2}=M.$ The scalar products of the simple roots are 
$\left( \pi _{2},\pi _{2}\right) =2,\left( \pi _{1},\pi _{1}\right) =1,\left(
\pi _{2},\pi _{1}\right) =-1$. If we consider now the spherical shell model,
where protons and neutrons can occupy single particle states with quantum
numbers $(j,m_{j})$ then the sub-index $\uparrow $ corresponds to $m_{j}>0$
and $\downarrow $ to $m_{j}<0$ and the single particle index $i$ labels the
states $(j,|m_{j}|).$ Due to the rotational symmetry, we can use the angular
momentum $j$ instead of $i$ but have to take into account the corresponding
degeneracy $\Omega _{j}=(2j+1)/2.$ Finally the weights $\Lambda _{i}^{a}$ are
the same for any $i$ and correspond to the fundamental representation of 
$\frak{so}(5)$, that is, $\Lambda ^{1}=0$, and $ \Lambda ^{2}=1.$ 
Putting all this together with the choice $\rho^{1}=0, \rho^{2}=-1/g$, and 
$z_{i}=2\varepsilon _{i}$ in the generalized Richardson equating (\ref{gReq})
we obtain the equations for the proton-neutron $T=1$ pairing that were given
also by Links et al \cite{Links-JPA35} and Asorey et al \cite{Asorey-et-al}: 
\begin{eqnarray}
\frac{1}{g} &=&\sum_{i=1}^{L}\frac{{{\Omega }_{i}}}{2{{\varepsilon }_{i}}-{\
v_{\alpha }}}+\sum_{\beta \neq \alpha }^{M}\frac{2}{{v_{\alpha }}-{v_{\beta }
}}+\sum_{\gamma =1}^{M-T}\frac{1}{{w_{\gamma }}-{v_{\alpha }}}
\label{Links eqs.} \\
0 &=&\sum_{\alpha =1}^{M}\frac{1}{{v_{\alpha }}-{w_{\gamma }}}+\sum_{\delta
\neq \gamma }^{M-T}\frac{1}{{w_{\gamma }}-{w_{\delta }}}, \quad
E =\sum_{\alpha =1}^{M}{v_{\alpha }}. \nonumber
\end{eqnarray}
The spectral parameters $v_{\alpha }$ have the same meaning as pair energies. 
If one allows  for isospin breaking then one has to set $\rho^{1}=\Delta\ne0$ 
\cite{so5pairing}.

\subsection{Spin-Isospin pn-pairing as $ \frak{so}(8)$ RG-model}\label{so8model}

In the previous section on our discussion of the T=1 proton-neutron pairing as 
$\frak{so}(5)$ RG-model, we considered the intrinsic symmetry space to be the isospin $SU_{T}(2)$ symmetry while the extrinsic spaces labeled by $i$ were related to the total spin states $(j,m_{j}).$ In particular, the pairs operators $\{b(\mu), b^{+}(\mu)\}$ and the $\frak{su}_{T}(2)$ algebra generators $\{T_{0},T_{\pm}\}$ were time-reversal invariant. For the study of the spin-isospin pn-pairing it is appropriate to consider the Wigner's $SU_{ST}(4)=SU_{S}(2)\times SU_{T}(2)$ as intrinsic symmetry of the system \cite{Wigner} and the orbital angular momentum  $(l,m_{l})$ as the extrinsic space labeled with $i$ in the appropriate sums. This way the protons and neutrons are described by the operators: ${a_{l_{i},m;s, \sigma;t,\tau}}$ where $s=t=\frac{1}{2}$ or briefly ${a_{l_{i},m,\sigma,\tau}}$. The pair operators are defined as isovector and spinvector tensor operators:
\[
P_{\tau i}=\sqrt{\Omega_{l_{i}}}\left[{a_{l_{i}}}{a_{l_{i}}}\right]^{001}_{00\tau},\quad
D_{\sigma i}=\sqrt{\Omega_{l_{i}}}\left[{a_{l_{i}}}{a_{l_{i}}}\right]^{010}_{0\sigma 0}.\\
\]
The $\Omega_{l_{i}}=(2l_{i}+1)/2)$ appears here due to the structure of the Clebsch-Gordan coefficients $<lm,l-m|00>$. The Wigner's $\frak{su}(4)$ along with the $\frak{u}(1)$ number operator 
$N=N_{n_{\uparrow}}+N_{n_{\downarrow}}
+N_{p_{\uparrow}} + N_{p_{\downarrow}}
=2n$
which is double of the pair number operator $n$, form $u(4)$ algebra with the following 16 generators:
\[X_{\tau_{1} \sigma_{1} \tau_{2} \sigma_{2} i}=\sum_{m}
a^{+}_{l_{i}m,\tau_{1} \sigma_{1}} 
a_{l_{i}m,\tau_{2} \sigma_{2}},
\]
that are $\frak{u}(1)$,  $\frak{su}_S(2)$, and $\frak{su}_T(2)$ tensors up to a factor $\sqrt{\Omega_{l_{i}}}$:
\[
N_{i}\sim\left[a^{+}_{l_{i}} a_{l_{i}} \right]^{000}_{000},
S_{\sigma i}\sim\left[a^{+}_{l_{i}} a_{l_{i}} \right]^{010}_{0\sigma0},
T_{\tau i}\sim\left[a^{+}_{l_{i}} a_{l_{i}} \right]^{001}_{00\tau},
Y_{\sigma\tau i}\sim\left[a^{+}_{l_{i}} a_{l_{i}} \right]^{011}_{0\sigma\tau}.
\]
The relevant hamiltonian has equal spin and isospin pairing strength:
\[
H_{P} =\sum_{i}^{L}2\varepsilon _{i} n_{i}-
g \sum_{ij,\mu}(P^{\dagger}_{\mu i}P_{\mu j}+D^{\dagger}_{\mu i}D_{\mu j})
\]
with eigenvalues:
$ E=\sum_{\alpha}^{M_1}{e_{\alpha}} $
where each $M_1$ pair contributes pair energies $e_{\alpha}$ determined by the four equations \cite{so8pairing}:

\begin{eqnarray*}
\frac{1}{g}&=&
\sum_{i}^L\frac{\Omega_{i}}{2 \epsilon_i-e_{\alpha}}
-\sum_{\alpha'(\not=\alpha)}^{M_1}\frac{2}{e_{\alpha'}-e_{\alpha}}
+\sum_{\alpha'}^{M_2}\frac{1}{\omega_{\alpha'}-e_{\alpha}}\\
0&=&
-\sum_{\alpha'}^{M_1}\frac{1}{e_{\alpha'}-\omega_{\alpha}}
+\sum_{\alpha'(\not=\alpha)}^{M_2}\frac{2}{\omega_{\alpha'}-\omega_{\alpha}}
-\sum_{\alpha'}^{M_3}\frac{1}{\eta_{\alpha'}-\omega_{\alpha}}
-\sum_{\alpha'}^{M_4}\frac{1}{\gamma_{\alpha'}-\omega_{\alpha}}\\
0&=&
-\sum_{\alpha'}^{M_2}\frac{1}{\omega_{\alpha'}-\eta_{\alpha}}
+\sum_{\alpha'(\not=\alpha)}^{M_3}\frac{2}{\eta_{\alpha'}-\eta_{\alpha}}
\nonumber\\
0&=&
-\sum_{\alpha'}^{M_2}\frac{1}{\omega_{\alpha'}-\gamma_{\alpha}}
+\sum_{\alpha('\not=\alpha)}^{M_4}\frac{2}{\gamma_{\alpha'}-\gamma_{\alpha}}
\nonumber.
\end{eqnarray*}
Since this is $ \frak{so}(8)$ RG model of rank is 4 there are 4 sets of spectral parameters. The number of spectral parameters in each set is determined by the relevant $u(4)$ Wigner multiplet. For a given number of pairs $M$, these  $u(4)$ multiplets can be classified using Young tableaux. Each multiplet  is defined by a partition of $M$ in  4 numbers,  $[\lambda_1 \lambda_2 \lambda_3 \lambda_4]$, constrained by: 
$ \sum_{i} \Omega_{i} \geq  \lambda_1\geq  \lambda_2\geq  \lambda_3\geq  \lambda_4\geq 0$.  
The labels $ \lambda_i$ are related to the number of pairs in the lowest/highest weight state.
Thus $M_1= \lambda_1+ \lambda_2+ \lambda_3+ \lambda_4$ is the number of pairs $M$, 
$M_2= \lambda_2+ \lambda_{3}+ \lambda_{4}$, $M_3= \lambda_3 +  \lambda_{4}$, and $M_4= \lambda_4$.

\section{Conclusions and Discussions}\label{nuclearmodels}

Using the Cartan classification of the semi-simple Lie algebras one can see
that many physics models can be generalized within the above framework.
In Table \ref{Table1} are shown the semi-simple Lie groups up to rank four
and the corresponding main fermion models. R\&R denotes the model developed 
by G. Rosensteel and D. J. Rowe and TCF stands for Trapped Cold Fermions \cite{colorPairing}.

\begin{table}[htbp]
\scriptsize
\begin{center}
\begin{tabular}{||c|c|c|c|c||}
\hline\hline
rank $n$ & $A_{n}\quad \frak{su}(n+1)$ & $B_{n}\quad \frak{so}(2n+1)$ & $C_{n}\quad \frak{sp}(2n) $ & $D_{n}\quad \frak{so}(2n)$ \\ 
\hline 1 & \textbf{$\frak{su}(2)$ pairing} & $\frak{so}(3)\sim \frak{su}(2)$ & $\frak{sp}(2)\sim \frak{su}(2)$ & $\frak{so}(2)\sim \frak{u}(1)$ \\ 
\hline 2 & $\frak{su}(3)$ Elliott & \textbf{$\frak{so}(5)$ T=1 pairing} & \textbf{$\frak{sp}(4)\sim \frak{so}(5)$} & $\frak{so}(4)\sim \frak{su}(2)\oplus \frak{su}(2)$ \\ 
\hline 3 & $\frak{su}(4)$ Wigner & $\frak{so}(7)\subset \frak{so}(8)$ FDSM & $\frak{sp}(6)$ R\&R
& \textbf{$\frak{so}(6)\sim \frak{su}(4)$ TCF}\\ 
\hline 4 & $\frak{su}(5)$ & $\frak{so}(9)$ & $\frak{sp}(8)$ & \textbf{$\frak{so}(8)$ Evans, FDSM} \\ 
\hline\hline
\end{tabular}
\caption{Group structure associated with important nuclear physics models; 
pairing models are in boldface. Isomorphisms of Lie algebras are denoted with $\sim$. }
\end{center}
\label{Table1}
\end{table}

The RG-models are a new important mathematical tool to study the behavior of physical systems.
They are finding applications to the such fields of studies as super-conducting grains, atomic nuclei, and trapped fermion atoms.

\section*{Acknowledgements}

V. Gueorguiev is grateful to his colleagues from the Bulgarian Academy of Sciences for the moral support and scientific encouragement, for their interest in his research, and for the many opportunities over the years to attend and present his research at their regular scientific meetings that they run very successfully over the years despite of the difficult economic times.


\end{document}